\documentclass[]{svmult}

\usepackage{makeidx}         
\usepackage{graphicx}        
\usepackage{multicol}        
\usepackage[bottom]{footmisc}

\makeindex             

\begin{document}

\title*{General Relativistic MHD Jets}

\author{Julian H. Krolik\inst{1}\and
John F. Hawley\inst{2}}
\institute{Johns Hopkins University, Baltimore MD USA
\texttt{jhk@jhu.edu}
\and University of Virginia, Charlottesville VA USA \texttt{jh8h@virginia.edu}}
\maketitle

\abstract{Magnetic fields connecting the immediate environs of rotating
black holes to large distances appear to be the most promising mechanism
for launching relativistic jets, an idea first developed by Blandford and
Znajek in the mid-1970s.  To enable an understanding of this process, we
provide a brief introduction to dynamics and electromagnetism in the spacetime
near black holes.   We then present a brief summary of the classical
Blandford-Znajek mechanism and its conceptual foundations.   Recently,
it has become possible to study these effects in much greater detail using
numerical simulation.   After discussing which aspects of the problem can
be handled well by numerical means and which aspects remain beyond the grasp
of such methods, we summarize their results so far.  Simulations have confirmed
that processes akin to the classical Blandford-Znajek mechanism can launch
powerful electromagnetically-dominated jets, and have shown how the jet luminosity
can be related to black hole spin and concurrent accretion rate.  However,
they have also shown that the luminosity and variability of jets can depend
strongly on magnetic field geometry.  We close with a discussion of several
important open questions.}

\section{The Black Hole Connection}
\label{sec:bhconnection}

     Jets can be found in association with many astronomical objects:
everything from Saturn's moon Enceladus to proto-stars to symbiotic
stars to planetary nebulae to neutron stars and the subject of this
chapter, black holes.  What distinguishes the black hole jets,
whether they are attached to stellar mass black holes or supermassive
black holes, is that they are the only ones whose velocities are relativistic.
That this should be so is not terribly surprising because, of course,
the immediate vicinity of a black hole is the most thoroughly relativistic
environment one could imagine.

     Although, as detailed elsewhere in this book, black holes can generate
jets in a wide variety of circumstances, the fundamental physics of black
holes can be treated in a unified manner.   The reason for this simplicity
is that the mass of the black hole changes the physical length scale for
events in its neighborhood, but little else.  Once distance is measured
in gravitational radii---$r_g \equiv GM/c^2 \simeq 1.5 (M/M_{\odot})$~km---the
mass becomes very nearly irrelevant.   Thus, jets from black holes
of roughly stellar mass (whether in mass-transfer binaries or in
the sources of $\gamma$-ray bursts) can be expected to behave in a way
very similar to those ejected from quasars, whose black hole mass can
be $10^8$ times larger.

      Another unifying theme to the physics of jets from black holes is
the central role of magnetohydrodynamics (MHD\index{MHD}).   In almost any conceivable
circumstances near black holes, the matter must be sufficiently ionized
to make it an excellent conductor.  When that is the case, the fluid
and the magnetic field cannot move across each other.  Because magnetic
fields are able to link widely-separated locations, they can provide the
large-scale structural backbones for coherent structures like jets.

\section{General Relativity Review}
\label{sec:gr}

     To properly understand the mechanics of matter and electromagnetic
fields near black holes,
it is, of course, essential to describe them in the language of relativity.
This is, therefore, an appropriate point at which to insert a brief
review of the portions of this subject most necessary to understand
how black holes can drive jets.

\subsection{Kinematics}
\label{sec:grkinematics}

     One of the fundamental tenets of general relativity is that mass-energy
induces intrinsic curvature in nearby space-time; what we call gravity
is the result.   Thus, in order to describe the motion of anything under gravity,
it is necessary to find the space-time metric, the relationship that determines
differential proper time $ds$, the time as it is perceived by a particle
in its own rest-frame, in terms of a differential separation in four-dimensional
space-time $dx^\mu$:
\begin{equation}
ds^2 = g_{\mu\nu}dx^\mu dx^\nu .
\end{equation}
By standard convention, all Greek indices vary over the integers 0,1,2,3,
and the zero-th index is the one most closely associated with time,
with the other three associated (more or less) with the usual three
spatial dimensions.  We will also adopt the (very common, but not
universal) convention that the metric signature is $-+++$.
Proper time is a physically well-defined quantity that is invariant to
frame-transformations; in contrast, motion in space-time can
be parameterized in terms of all sorts of different coordinate systems
that, on their own, do not necessarily have to possess any sort of
physical significance.

     Of the many possible ways to erect coordinate systems around
black holes, two are most commonly used: Boyer-Lindquist coordinates\index{Boyer-Lindquist coordinates}
and Kerr-Schild coordinates\index{Kerr-Schild coordinates}.   Both are built on conventional spherical
coordinates for the spatial dimensions.   The former has the advantage
that its coordinate time is identical to the proper time of an observer at
large distance from the point mass, but the disadvantage that some of
the elements in the metric diverge at the black hole's event horizon.
The latter has the advantage of no divergences, but the disadvantage
that its time coordinate bears a more complicated relation to time
as seen at infinity.

      In Boyer-Lindquist coordinates, the metric around a rotating
black hole is
\begin{eqnarray}
ds^2 = &-&\left( 1- {2Mr \over \Sigma}\right)dt^2 - {4aMr\sin^2\theta \over
\Sigma} dt d\phi + {\Sigma \over \Delta} dr^2 \\
\quad  &+& \Sigma d\theta^2 + \left(r^2 + a^2 + {2Mra^2\sin^2\theta \over
 \Sigma}\right) \sin^2\theta d\phi^2 ,\\
\end{eqnarray}
where
\begin{eqnarray}
\Sigma &\equiv& r^2 + a^2 \cos^2\theta \\
\Delta &\equiv& r^2 - 2Mr + a^2 , \\
\end{eqnarray}
and the polar direction of the coordinates coincides with the direction
of the angular momentum.  The ``spin parameter" $a$ is defined such
that the black hole's angular momentum $J = aM$.   Here, and in all
subsequent relativistic expressions, we take $G=c=1$.

       When $r/M \gg 1$, the Boyer-Lindquist coordinate system clearly
reduces to that of spherical spatial coordinates in flat, empty space.
On the other hand, when $r/M \sim O(1)$, there are odd-looking complications,
most notably a non-zero coupling, proportional to $a$, between motion in
the azimuthal direction $\phi$ and the passage of time.   That there
is such a coupling suggests that particles must rotate around the black
hole (i.e., change $\phi$ over time) because of the properties of the
space-time itself, rather than because they have intrinsic angular
momentum.   The truth of this hint is seen clearly after a transformation
to a closely-related system of coordinates.  In this new system,
$r$, $\theta$, and $t$ are the same as in Boyer-Lindquist coordinates\index{Boyer-Lindquist coordinates},
but the azimuthal position is described in terms of what would be
seen by an observer following a circular orbit in the plane of the
black hole's rotation with frequency $\omega_c = -g_{t\phi}/g_{\phi\phi}.$
The relation between the new azimuthal coordinate and the old is
$d\phi^\prime = d\phi - \omega_c dt$, and the resulting metric is
\begin{equation}
ds^2 = -{\Sigma \Delta \over A} dt^2 + {A \over \Sigma} \sin^2\theta
\left(d\phi^{\prime}\right)^2 + {\Sigma \over \Delta} dr^2
+ \Sigma d\theta^2,
\end{equation}
where the new function $A$ is
\begin{equation}
A = (r^2 + a^2)^2 -a^2 \Delta \sin^2\theta.
\end{equation}
This metric is diagonal, so relative to its coordinates there
is no required rotation.
For this reason, it is sometimes called the ``LNRF" (for ``Locally
Non-Rotating Frame\index{Locally
Non-Rotating Frame}") coordinate system.  It is also sometimes called
the ``ZAMO" (for ``Zero Angular Momentum") system because (as one
can easily show), particles moving in the equatorial plane with zero
angular momentum follow trajectories with $d\phi^\prime = 0$.   In other
words, near a black hole, even zero angular momentum orbits must
rotate relative to an azimuthal coordinate system fixed with respect
to distant observers.  Another consequence of the metric's diagonality
is that it is easy to read off the lapse function, the gravitational
redshift factor: it is $\alpha = \sqrt{\Sigma \Delta/A}$.

      Further physical implications of this metric can be found
by studying the behavior of the four-velocity $u^\mu \equiv dx^\mu /ds$
for any physical particle with non-zero rest-mass.  As a consequence
of the Equivalence Principle, particles travel along geodesics, paths
that maximize accumulated proper time.  The metric can then be viewed
as akin to a Lagrangian and framed in terms of canonical coordinates
and their conjugate momenta.  Because $dx^\mu = u^\mu ds$, the covariant
components of the four-velocity effectively become the momenta conjugate
to the coordinates.  In particular, when the metric is time-steady and
axisymmetric about an axis, $u_t$ is the conserved orbital energy and
$u_\phi$ is the conserved angular momentum for motion about that axis.
This conserved orbital energy is often called the energy-at-infinity
($E_{\infty}$)---after all, if the orbit extends to infinity, that is
still the energy it would have there.  Consider, for example, a particle
in a circular orbit with coordinate orbital frequency $\Omega$; that is,
$\Omega = u^\phi/u^t$.  With the minus sign demanded by our sign convention,
its conserved energy is
\begin{equation}
E_{\infty} = -u_t = -\left(g_{tt}u^t + g_{t\phi}u^\phi\right) = 
               -u^t\left(g_{tt} + g_{t\phi}\Omega\right). 
\end{equation}
Ordinarily, $E_{\infty} > 0$ because $g_{tt} < 0$ and $g_{t\phi}$ is small.
However, close to a rapidly-rotating black hole, these sign relations
can change: $g_{tt}$ becomes negative when $r < 2M$ (in the equatorial
plane) and $g_{t\phi}$ can be $\sim O(1)$.   This region is called
the ``ergosphere".  Inside the ergosphere, rotations that are relatively
large and retrograde can lead to $-u_t < 0$.  That
is, the orbital energy, even including rest-mass, can become negative!

     The event horizon is found at smaller radius than the edge of
the ergosphere when the black hole possesses any spin.  It forms
the constant-$r$ surface $r/M = 1 + \sqrt{1 - (a/M)^2}$.  On that surface,
$\Delta = 0$, so $g_{rr}$ (in Boyer-Lindquist coordinates\index{Boyer-Lindquist coordinates})
diverges.   As we shall see momentarily, this is only a formal
divergence and indicates nothing odd physically (or at least
nothing stranger than an event horizon!).

     Alternatively, motions around a rotating point-mass can be
described in terms of Kerr-Schild coordinates.   These differ
from Boyer-Lindquist by the coordinate transformation
\begin{eqnarray}
dt_{KS} &=& dt_{BL} + {2Mr \over \Delta}dr \\
dr_{KS} &=& dr_{BL}\\
d\phi_{KS} &=& d\phi_{BL} + {a \over \Delta}dr\\
d\theta_{KS} &=& d\theta_{BL}.\\
\end{eqnarray}
The corresponding metric is
\begin{eqnarray}
\hskip -0.9pc 
ds^2 = -&(&1 - \beta_r)dt^2 + 2\beta_r dt dr+ (1 + \beta_r)dr^2 + \Sigma d\theta^2
- 2a\beta_r\sin^2\theta dt d\phi\\
 &+& \left(r^2+a^2 + {2Mra^2\sin^2\theta \over \Sigma}\right) \sin^2\theta d\phi^2 
  -   2a(1 + \beta_r)\sin^2\theta dr d\phi. \\
\end{eqnarray}
Here $\beta_r = 2Mr/\Sigma$.
Note that there is no divergence in any of the metric elements at the
event horizon or anywhere else (except the origin).  Because physical
results cannot depend
on our choice of coordinates, this fact demonstrates that the Boyer-Lindquist
divergence at the horizon is purely an artifact of that coordinate system.
On the other hand, the same divergence occurs in the relation between
these two coordinate systems.  The closer to the event horizon one probes,
the faster Kerr-Schild time advances relative to Boyer-Lindquist time.
Similarly, close to the event horizon, the Kerr-Schild azimuthal angle
advances rapidly relative to the Boyer-Lindquist azimuthal angle.
Thus, the convenience the Kerr-Schild system provides in avoiding divergences
is partially offset by a price paid in ease of physical interpretation.

   The last comment worth making here is that often the most interesting
frame in which to evaluate quantities is that of the local
fluid motion itself.  To do so most conveniently, one erects a local
system of orthonormal unit vectors called a ``tetrad".   By standard
convention, the tetrad element pointing in the local time direction is
$\hat e^\mu_{(t)} \equiv -u^\mu$ (the minus sign is the result of
our choice of metric signature).  In the fluid frame, the only non-zero
component of the four-velocity is the rate of advance of proper time
($=1$, of course), so this makes a natural definition of the local
sense of time.  There is considerable freedom in how the spatial
tetrads are oriented, but it is always possible to construct a complete
set via a Gram-Schmidt procedure.  When such a tetrad is available,
any tensor quantity may be evaluated as it
would be measured in the fluid frame.  All that is required is to project
onto the appropriate tetrads: for example, a four-vector $X^\mu$ as seen
in the fluid frame is $X^\nu \hat e_\nu^{(\mu)}$.

\subsection{Electromagnetic Fields}

     One way to develop an appropriately covariant formulation of
electromagnetism is to begin with a 4-vector potential $A_\mu$.  Its
elements have the usual interpretation: the time-component is related
to the electrostatic potential, while its spatial components determine
the 3-vector whose curl is the magnetic field.  Thus, we can construct
the Maxwell tensor
\begin{equation}
F_{\mu\nu} = \nabla_{\mu}A_\nu - \nabla_{\nu}A_\mu,
\end{equation}
so that what we might call the electric field ${\cal E}^i = F_{ti}$ and
the magnetic field is ${\cal B}^i = [ijk]F_{jk}$.   Here $[ijk]$ is
the antisymmetric permutation operator and the symbol $\nabla_\mu$
denotes a covariant derivative, i.e., a derivative that accounts for
changes in the direction of local unit vectors as well as changes in
vector components.

To see the electric and
magnetic field components as elements of a tensor rather than as
vector quantities may seem in conflict with the usual way to think
about these fields.  However, the Maxwell tensor can be easily related
to a vector version of the fields through the construction
\begin{eqnarray}
E^\mu &=& u_\nu \, ^{*}F^{\mu\nu} \\
B^\mu &=& u_\nu \, ^{*}F^{\mu\nu},\\
\end{eqnarray}
where $^{*}F$ is the dual of $F$, i.e.,
\begin{equation}
^{*}F^{\mu\nu} = (1/2)[\alpha\beta\mu\nu]F_{\alpha\beta}. 
\end{equation}
In the rest frame of the fluid,
only the time component of $u_\mu$ is non-zero; thus, the spatial
parts of $E^\mu$ and $B^\mu$ may be interpreted as the electric and
magnetic fields as they appear in the fluid rest-frame.  Moreover,
because the field tensor is anti-symmetric, the time-component of
both $E^\mu$ and $B^\mu$ is always zero in the fluid frame.

     In this language, Maxwell's equations are simply
\begin{eqnarray}
&\nabla_{\mu}& F^{\mu\nu} = 4\pi J^\nu\\
&\nabla_{\mu}& *F^{\mu\nu} = 0.\\
\end{eqnarray}
Here the four-current density
\begin{equation}
J^\mu = qnu^\mu,
\end{equation}
for particle charge $q$, particle proper number density $n$, and four-velocity
$u^\mu$.

\subsection{Dynamics: the Stress-Energy Tensor}

Mechanics can be thought of as the conservation laws at work in the world.
To apply the laws of conservation of energy and momentum in an
electromagnetically-active relativistic
setting, we first define the stress-energy tensor
\begin{equation}
T^{\mu}_{\nu} = \rho h u^{\mu}u_{\nu} + pg^{\mu}_{\nu} + 
{1 \over 4\pi}\left[F^{\mu\alpha}F_{\nu\alpha}
- {1 \over 4}g^{\mu}_{\nu}F^{\alpha\beta}F_{\alpha\beta}\right],
\end{equation}
where $h = 1 + (\epsilon + p)/\rho$ is the relativistic enthalpy, $\rho$ is the
proper rest-mass density, $\epsilon$ is the proper internal energy density,
and $p$ is the proper pressure.   The first
two terms in this expression are manifestly the relativistic generalizations
of the flux of momentum.  The second two terms give the electromagnetic
contribution, which can also be written in a more familiar-appearing way:
\begin{eqnarray}
F^{\mu\alpha}F_{\nu\alpha} - 
   {1 \over 4}g^{\mu}_{\nu}F^{\alpha\beta}F_{\alpha\beta} =
\left(||E||^2 + ||B||^2 \right)&u^\mu& u_\nu -E^{\mu}E_{\nu} - B^{\mu} B_{\nu} \\
+ &{1 \over 2}&\left(||E||^2 + ||B||^2 \right)g^{\mu}_{\nu}.\\
\end{eqnarray}
In other words, the electric and magnetic energy densities contribute to:
the total energy density conveyed with the fluid (the term
$\propto u^\mu u_\nu$); the pressure (the final term); and the momentum flux
(the $E^\mu E_\nu$ and $B^\mu B_\nu$ terms).

When the characteristic rate at which the fields are seen to change in
the fluid rest-frame is slow compared to the electron plasma frequency
$(4\pi_e n e^2/m_e)^{1/2}$, the electrons can quickly respond to
imposed electric fields and cancel them.  Because the most rapid
conceivable rate at which anything can change is $\sim c/r_g$, this
criterion translates to $n_e \gg 14 (M/M_{\odot})^{-2}$~cm$^{-3}$.
If $E^\mu = 0$ in the fluid frame, it must be zero in all
frames.  In other words, the MHD condition, the limit in which charges
can flow freely and quickly in response to imposed fields, places
the constraint
\begin{equation}
E^\mu = u_\nu F^{\mu\nu} = 0.
\end{equation}
In the MHD limit, then, which ordinarily is very well-justified in the
vicinity of a black hole, energy-momentum conservation is expressed by
\begin{equation}
\nabla_{\mu}T^{\mu}_{\nu} = \nabla_{\mu}\left\{\rho h u^{\mu}u_{\nu} + 
\left(p + {||B||^2 \over 8\pi}\right)g^{\mu}_{\nu} + 
{1 \over 4\pi}\left[||B||^2 u^\mu u_\nu - B^{\mu}B_{\nu}\right]\right\} = 0.
\end{equation}

\section{Candidate Energy Sources for Jets}
\label{sec:energysources}

    Fundamentally, there are only two possible sources to tap for the
energy necessary to drive a jet: the potential energy released by accreting
matter and the rotational kinetic energy (more formally, the reducible
energy) of a rotating black hole.

    Consider the accretion energy first.  The net rate at which energy is
deposited in a ring of an accretion disk is the difference between the
divergences of two energy fluxes: Inter-ring stresses do work, thereby
transferring energy outward, while accreting matter brings its energy
inward.  However, there is an inherent mismatch between these two
divergences, which in a steady-state disk is always positive.  That is,
the energy deposited by the divergence of the work done by stress always
outweighs the diminution in energy due to the net inflow of binding
energy brought by accretion.  If disk dynamics alone are considered, there
are only two possible ways to achieve balance: by radiation losses or by
inward advection of the heat dissipated along with the accretion flow.
However, in principle the heat could also be used to drive an outflow,
e.g., a jet.

    These qualitative statements are readily translated to the
quantitative stress-tensor language just formulated.  In the simple
case of a time-steady and axisymmetric disk (viewed in the orbiting
frame),
\begin{eqnarray}
\int \, dz \left\{ \partial_r \left[\rho h u^r u_t 
+ \frac{1}{4\pi}\left(||B||^2 u^r u_t - B^r B_t\right)\right]
+ \partial_z T^z_t \right\} &=& 0\\
\partial_r \int \, dz \, \left[T^r_t({\rm matter}) + T^r_t({\rm EM})\right]
+ 2F &=& 0.\\
\end{eqnarray}
Here $T^z_t$ is nothing other than the vertical flux of energy, perhaps
in radiation, hence the natural identification of $T^z_t$ at the top
and bottom surfaces as the outward flux $F$.  Because that outward flux
cannot be negative in an isolated accretion flow, the net energy deposited
by work must always exceed the loss due to inflow.

     How this dissipation takes place is, of course, unspecified by
arguments based on conservation laws.  Given the preeminence of magnetic
forces in driving accretion, we can reasonably expect most of the
heat to be generated by dissipating magnetic field.  Reconnection
events and other examples of anomalous resistivity such as ion transit-time
damping, etc. are good candidate mechanisms, but little definite is known
about this subject (see, e.g., \cite{kuz07} and references therein).
Nonetheless, given the nature of all these mechanisms,
which are characteristically triggered by sharp gradients, it is very
likely that the heating is highly localized and intermittent.  For
this reason, it is quite plausible that the dissipation rate is sufficiently
concentrated as to drive small amounts of matter to temperatures comparable
to or greater than the local virial temperature.  If this is the case,
disk dissipation could substantially contribute to driving outflows.

      Disk heating may also help expel outflows through a different
mechanism: radiation forces.  As mentioned before, the energy of net
disk heating can be carried off by radiation.  Because the acceleration
due to radiation is $\kappa \vec{\cal F}/c$, wherever the opacity per
unit mass $\kappa$ is high enough, radiation-driven outward acceleration
may surpass the inward acceleration of gravity.  Although most other
elements of accretion dynamics onto black holes are insensitive to
the central mass, the opacity, through its dependence on temperature,
can depend strongly on it.  In particular, the opacity of disk matter
in the inner rings of AGN disks, where the temperature is only $\sim 10^5$~K,
may be several orders of magnitude greater than in the inner rings of
Galactic black hole disks, where the temperature is so high ($\sim 10^7$~K)
that almost all elements are thoroughly ionized.  It may consequently
be rather easier for AGN disks to expel radiation-driven winds than
for Galactic black holes, even though the accretion rate is still
well below Eddington.

      Black hole rotation may also power outflows.  The Second Law of
Black Hole Thermodynamics decrees that the area of a black hole cannot
decrease, but diminishing spin (at constant mass) {\it increases} the
area.  Consequently, a black hole of mass $M$ and spin parameter $a$
has a reducible mass, that is, an amount of mass-energy that can be
yielded to the outside world, of $1 - [(1+\sqrt{1-(a/M)^2})/2]^{1/2}$.

       Given the intuitive picture of a black hole as an object that
only accepts mass and energy, one might reasonably ask how one can
give up energy.  One way to recognize that this may be possible is
to observe that the rotational frame-dragging a spinning black hole
imposes on its vicinity permits it to do work on external matter,
and in that way lose energy.  Another way to see the same point is
to recall that it is possible for particles inside the ergosphere
to find themselves on negative energy orbits.  If they cross the
event horizon on such a trajectory, their negative contribution
to the black hole's mass-energy results in a net decrease of its
mass-energy, or a release of energy to the outside world, as
originally pointed out by \cite{p69}.  These negative
energy orbits in general involve retrograde rotation, and so
likewise bring negative angular momentum to the black hole.
Collisions between a pair of positive energy particles that
result in one of the two being put onto a negative energy orbit
and then captured by the black hole are called the ``Penrose
process", and are the archetypal mechanism for deriving energy
from a rotating black hole.   Perhaps regrettably, the kinematic
constraints for such collisions are so severe as to make them
extremely rare \cite{bpt72}.

      It is also possible for electromagnetic fields to bring
negative energy and angular momentum through the event horizon.
Consider, for example, an accretion flow that is in the MHD
limit, so that the magnetic field and the matter are tied together.
The Poynting flux at the event horizon can then be outward even
while the flow moves inward if the electromagnetic energy-at-infinity
is negative.  As shown by \cite{k05}, the density of this quantity can
be written (in Boyer-Lindquist coordinates\index{Boyer-Lindquist coordinates}) as
\begin{eqnarray}
e_{\infty,EM}/\alpha &=& -T^t_t (EM) \\
                 &=& -(1/2)g^{tt}\left[ g^{rr} ({\cal E}^r)^2
   + g^{\theta\theta}({\cal E}^\theta)^2 + g^{\phi\phi}({\cal E}^\phi)^2\right]
   + (1/2)\left(g^{t\phi} {\cal E}^\phi\right)^2 \\
   &+& \left[ g^{\theta\theta}g^{\phi\phi} ({\cal B}^r)^2
   + g^{rr}g^{\phi\phi}({\cal B}^\theta)^2 + 
      g^{rr}g^{\theta\theta}({\cal B}^\phi)^2\right].\\
\end{eqnarray}
Here $T^\mu_\nu(EM)$ is the electromagnetic part
of the stress-energy tensor, and ${\cal E}^i,{\cal B}^i$ are the
elements of the Maxwell tensor.  If all the spatial metric elements were
positive-definite, $e_{\infty,EM}$ would be likewise ($g^{tt} < 0$).
However, inside the ergosphere $g^{\phi\phi} < 0$.  Thus, wherever inside
the ergosphere the poloidal components of the magnetic field and the
toroidal component of the electric field are large, the electromagnetic
energy-at-infinity can become negative.

There is a close analogy between negative electromagnetic energy-at-infinity
and negative mechanical energy-at-infinity.  In the case of particle
orbits, the energy goes negative when the motion is in the ergosphere,
and is sufficiently rapidly retrograde with respect to the black hole
spin.   As \cite{kom08} has shown, electromagnetic waves have
negative energy-at-infinity when their normalized wave-vector as viewed
in the ZAMO frame has an azimuthal component $< -\alpha/\omega_c$.
This becomes possible only inside the ergosphere because it is only
there that $\alpha < \omega_c$.  In addition, this constraint demonstrates
that negative electromagnetic energy-at-infinity is also associated
with motion that is sufficiently rapidly retrograde.

\section{The Blandford-Znajek Mechanism}
\label{sec:b-z}

     As we have already seen, it is possible for black holes
to give up energy to the outside world both by swallowing
material particles of negative energy and by accepting negative
energy electromagnetic fields.  The latter mechanism can be much more
effective.  That this is so is largely due to the large-scale
connections that magnetic fields can provide, as first noticed
by \cite{bz77}.

    In that extremely influential paper, Blandford and Znajek
pointed out that a magnetic fieldline stretching from infinity
to deep inside a rotating black hole's ergosphere can readily
transport energy from the black hole outward.  For its
original formulation, the Blandford-Znajek mechanism\index{Blandford-Znajek mechanism} was
envisioned in an extremely simple way: a set of axisymmetric
purely poloidal fieldlines stretch from infinity to the edge
of the event horizon and back out to infinity, and are in a
stationary state of force-free equilibrium.  There is just
enough plasma to support the currents associated with this
magnetic field and to cancel the electric field in every local
fluid frame, but its inertia is entirely negligible.  These
assumptions were relaxed slightly in the work of \cite{p83},
who showed that inserting just enough inertia of matter
to distinguish the magnetosonic speed from $c$ does not materially
alter the result.

Given these assumptions, the Poynting flux on the black hole horizon
can be found very simply in terms of the magnitude of the radial
component of the magnetic field there and the rotation rate of the
field lines.  Because $E_\mu = 0$, the electromagnetic invariant
$\vec E \cdot \vec B = E_\mu B^\mu = 0$.  This, in turn, implies
that $^{*}F^{\mu\nu} F_{\mu\nu} = 0$.  When the fields are both
axisymmetric and time-steady, the product of the Maxwell tensors
reduces to a single identity
\begin{equation}\label{BZident}
\left(\partial_\theta A_{\phi}\right)\left(\partial_r A_{t}\right) =
\left(\partial_\theta A_t\right)\left(\partial_r A_\phi\right).
\end{equation}
Also because of the assumed stationarity and axisymmetry,
\begin{eqnarray}
\partial_r A_t &=& {\cal E}^r \\
\partial_\theta A_\phi &=& {\cal B}^r \\
\partial_\theta A_t &=& {\cal E}^\theta \\
\partial_r A_\phi &=& {\cal B}^\theta .\\
\end{eqnarray}
Consequently, the identity of equation~\ref{BZident} may be rewritten as
\begin{equation}
{{\cal E}^\theta \over {\cal B}^r} = {{\cal E}^r \over {\cal B^\theta}}.
\end{equation}
Because rotation through a magnetic field creates an electric field,
both ratios in the previous equation can be interpreted as a rotation
rate $\Omega_F$ associated with the poloidal field lines.

With all these relations in hand, it is straightforward to evaluate the
electromagnetic energy flux, i.e., the electromagnetic part of $-T^r_t$.
As \cite{mg04} pointed out, the algebra to do this is more concise in
Kerr-Schild coordinates than in Boyer-Lindquist, and the result is:
\begin{equation}
{\cal F} = -\Omega_F\sin^2\theta \left[2({\cal B}^r)^2 r\left(\Omega_F -
                                                           \frac{a/M}{2r}\right)
                      + {\cal B}^r{\cal B}^\phi \left(r^2 - 2r + (a/M)^2\right)\right]
\end{equation}
At the horizon, $r^2 - 2r + (a/M)^2 = 0$, and the rotation rate of the black hole
itself $\Omega_H = (a/M)/(2r)$, so
\begin{equation}
{\cal F}_H = -2\Omega_F\sin^2\theta({\cal B}^r)^2 r\left(\Omega_F - \Omega_H\right).
\end{equation}
Note that outgoing flux depends critically on the field lines rotating
more slowly than the black hole.  Because spacetime itself immediately
outside the horizon rotates at $\Omega_H$ (as viewed by a distant observer),
if there is some load that always keeps the field lines moving
more slowly, there is a consistent stress through which the black hole does
work on the field.

This formula neatly describes the Poynting flux in terms of the
strength of the magnetic field and its rotation rate, but on its
own it cannot tell us the luminosity of the system because both the
strength of the field itself and the field line rotation rate are
left entirely undetermined.  In \cite{bz77},
${\cal B}^r$ and ${\cal B}^\phi$ are found in terms of a field
strength at infinity by assuming a specific field configuration
(split monopole or paraboloidal) and performing an expansion in
small $a/M$.  Separately, MacDonald and Thorne \cite{mt82} argued
that the power generated was maximized when $\Omega_F = (1/2)\Omega_H$,
where $\Omega_H$ is the rotation rate of the black hole itself.  However,
the Blandford-Znajek model {\it per se} has no ability to determine either the
magnitude of the field intensity or the field line rotation rate.

Raising a specific version of the general question about how rotating
black holes can give up energy, Punsly and Coroniti \cite{p90} questioned
whether this mechanism can operate on the
ground that no causal signal can travel outward from the event
horizon.  A summary answer to their question is that the actual conditions
determining outward energy flow are determined well outside the
event horizon, and that accretion of negative energy is possible
when the black hole rotates.  In addition, only rotating black holes have
reducible mass that can be lost.  This summary can be elaborated
from several points of view, all of which are equivalent.

One, which we have already mentioned,
but was not explored in the original Blandford-Znajek paper,
is that EM fields deep in the ergosphere can be driven to
negative energy by radiating Alfven waves outward.  Accretion
of those negative EM energy regions then amounts to an outward
Poynting flux on the event horizon.  In the special case of stationary flow,
a critical surface outside the event horizon can be
found within which information travels only inward.  The inward
flux of negative electromagnetic energy may then be considered
to be determined at this critical point \cite{t90,bz00}.

Another, which can be found in \cite{tpm86}, is
to note that the enforced rotation of fieldlines within the
ergosphere creates an electric field.  This electric field
can in turn drive currents that carry usable energy off to
infinity.  Indeed, \cite{kom08} argues that such a poloidal
current is part-and-parcel of the plasma's electric field
screening.

A third way to look at this same process is that frame-dragging
forces fieldlines to rotate that would otherwise be purely radial.
As a result, toroidal field components are created---and transverse
field is the prerequisite for Poynting flux.

Although the Punsly-Coroniti question has now been put to rest,
there are a number of other questions left unanswered by the classical
form of the Blandford-Znajek model.  Because it is a time-steady solution,
by definition it does not consider how the field got to its equilibrium
configuration.  One might then ask, in the context of trying to understand
why certain black holes support strong jets and others don't, whether
{\it intrinsic} large-scale field is a prerequisite for jet formation,
or whether field structures contained initially within the accretion
flow can expand to provide this large-scale field framework.  Another
question is whether the force-free (or nearly force-free) assumption,
while surely valid in much of the jet, yields a complete solution: for
example, in the equatorial plane of the accretion flow, one would generally
expect a breakdown in this condition; could that affect the global
character of the jet generated?  Still another question would be how
we can extend this model quantitatively to more general field shapes
and higher spin parameters.  Lastly, if field lines threading the
event horizon can carry Poynting flux to infinity, perhaps they
can also carry Poynting flux to the much nearer accretion disk: is
there an interaction between Blandford-Znajek-like behavior and
accretion?

As we shall see, explicit MHD simulations make these approximations and
limitations
unnecessary and allow us to answer (or at least begin to answer) many of
the questions left open by the original form of the Blandford-Znajek
idea.  These numerical calculations explicitly find the magnetic field
at the horizon on the basis of the field brought to the black hole by the
accretion flow, as well as the coefficients that replace $\Omega_F$ when, because
of time-variability and a breakdown of axisymmetry, it is no longer possible
to give that quantity a clear definition.  They can also determine the
shape of the field, its connection to the disk, etc., and work just as
well when $a/M$ approaches unity as when it approaches zero.

\section{What Simulations Can and Cannot Do}
\label{sec:simscant}

Before presenting the results of jet-launching simulations, it is worthwhile
first to put them in proper context by explaining which questions
they can answer well, and which not so well.  Some of the considerations
governing these distinctions are built into the very nature of
numerical simulations, but others merely reflect the limitations of
the current state-of-the-art.

First and foremost, simulation codes are devices for solving algebraically
complicated, nonlinear, coupled partial differential equations.  After
discretization, these equations can be solved by a variety of numerical
algorithms designed so that, when applied properly, the solutions that are
found converge to the correct continuous solution as the discretization is
made finer and finer.  Analytic
methods are far weaker at solving problems of this kind, particularly
those with strong nonlinearities.  The ability to cope with nonlinearity
makes numerical methods especially advantageous for studying
problems involving strong turbulence (an essential ingredient of
accretion disks), as turbulence is fundamentally nonlinear.
In addition, algorithms can be devised that maintain important
constraints (e.g., $\nabla \cdot \vec B = 0$, energy and momentum conservation,
etc.) to machine accuracy.  Employing these built-in constraints, most
jet/accretion codes are very good, for example, at conserving angular momentum
and using the induction equation to follow the time-dependence of the magnetic field.

A brief discussion of the two principal varieties of code that have been
employed to date will serve to illustrate how their methods achieve these
ends.  One such family (exemplified by the Hawley-De Villiers general
relativistic MHD\index{MHD} code {\it GRMHD} \cite{dVh03}) derives from the {\it Zeus}
code \cite{sn92}.
In these codes, hyperbolic partial differential equations that are first-order
in time but of arbitrary order in space are written as finite difference
equations of the form
\begin{equation}
U_i^{k+1}(\mathbf{x}) = U_i^{k}(\mathbf{x}) + 
   \left[S_i^k(\mathbf{x}) + T_i^k(\mathbf{x})\right]\Delta t,
\end{equation}
where the quantity $U_i^k$ is one of the dependent variables (velocity,
density, magnetic field , etc.) at the $k$-th time-step at spatial
grid-point $\mathbf{x}$, and $\Delta t$ is the length of the time-step.
The various terms that may enter into defining
the time-derivative are divided into ``source" terms $S_i$, terms that
are local in some sense (e.g., the pressure gradient) and ``transport"
terms $T_i$, terms that describe advection (e.g., the terms describing
the passive transport of momentum or energy with the flow).  The source
and transport terms are generally handled separately.  When conserved
quantities are carried in the transport terms, time-centered differencing
can improve the fidelity with which they are conserved.  Organizing the
grid so that the dominant velocity component is along a grid axis also
improves the quality of conservation for the momentum component in that
direction.  Although not strictly speaking a requirement of this method,
it is most often implemented with an energy equation that follows only
the thermal energy of the gas, ignoring any interchange between that
energy reservoir and the orbital and magnetic energy, except as required
by shocks.  Coherent motions automatically conserve orbital
energy through the conservation of momentum, but this approximation
ignores losses of kinetic and magnetic energy that occur as a result
of gridscale numerical dissipation.  The reason for this choice is
that in many cases the thermal energy is so small compared to the
orbital energy that it would be ill-defined numerically if a total
energy equation were solved and the thermal energy only found later
by subtracting the other, much larger, contributions to the total.

The other family organizes the equations differently.  In this approach
(e.g., the general relativistic MHD codes {\it HARM} \cite{gmt03} and
{\it HARM3D} \cite{nk08}),
conservation laws are automatically obeyed precisely because
the equations of motion are written in conservation form, i.e.,
\begin{equation}
\frac{\partial U_i}{\partial t} = -\nabla \cdot \mathbf{F_i} + S_i,
\end{equation}
with $U_i$ a density, $\mathbf{F_i}$ the corresponding flux, and $S_i$
again the source term.  Individual Riemann problems are solved across
each cell boundary in order to guarantee conservation of all the quantities
that should be conserved.  However, because the conserved densities and
fluxes are often defined in terms of underlying ``primitive variables"
(e.g., momentum density is $\rho \vec v$), after the time-advance one
must solve a set of nonlinear algebraic equations to find the new
primitive variables implied by the new conserved densities.  Clearly,
in this method, it is hoped that the advantages of conserving total energy
outweigh the disadvantages of local thermal energies that may be
subject to large numerical error.

In both styles, an initially divergence-free magnetic field can be
maintained in that condition by an artful solution of the induction
equation called the ``constrained transport" or CT method \cite{eh88,t00}.
The essential idea behind this scheme is to rewrite the differenced
form of the induction equation as an integral equation over each cell,
and then use Stokes' Theorem to transform cell-face integrals of
$\nabla \times (\mathbf{v} \times \mathbf{B})$ into cell-edge integrals
of $\mathbf{v} \times \mathbf{B}$.  Because the latter can be done
exactly, the induction equation itself can be solved exactly, and
the divergence-free condition preserved.

Unfortunately, however, this list of the strengths of numerical methods
does not encompass all possible problems that arise in black hole jet
studies.  Contemporary algorithms are particularly weak in those aspects
involving thermodynamics.  Two large gaps in our knowledge, one about
algorithms, the other about physics, make this a very difficult
subject: First, as in most areas of astrophysics, temperature regulation is the
result of photon emission, but there are as yet no methods for solving
3-d time-dependent radiation transfer problems quickly enough that they
would not drastically slow down a dynamical simulation code.  Second,
although the dense and optically thick conditions inside most accretion
disks make it very plausible that all particle distribution functions
are very close to thermodynamic equilibrium, outside disks, whether
in their coronae or, even more so, in their associated jets, this assumption
is, to put it mildly, highly questionable.  We would need a far better
understanding of plasma microphysics to improve upon this situation.
The combination of these two gaps makes it very hard to determine
reliably the local pressure (whether due to thermalized atoms,
non-thermal particles, or radiation), and associated hydrodynamic
forces.  As a result, the most reliable results of these calculations
have to do with dynamics for which gravity and magnetic forces dominate
pressure gradients.

Limitations in computing power create the next set of stumbling blocks.
The nonlinearities of turbulence may be thought of as transferring
energy from motions on one length scale to motions on another.  Generally
speaking, turbulence is stirred on comparatively long scales, these
nonlinear interactions move the energy to much finer scale motions,
and a variety of kinetic mechanisms, generically increasing in power
as the scale of variation diminishes, dissipate the energy into heat.
Unfortunately, the computer time required in order to run a given
3-d simulation with spatial resolution better by a factor of $R$
scales as $R^4$ (three powers from the increased number of spatial
cells, one power from the tighter numerical stability limit on the
size of the timestep).  Consequently, they are generally severely
limited in the dynamic range they may describe between the long
stirring length scale and the shortest scale describable, the gridscale.
The gridscale is therefore almost always many many orders of
magnitude larger than the physical dissipation scale.

If MHD turbulence in accretion behaves like hydrodynamic Kolmogorov
turbulence in the sense that it develops an ``inertial range" in which
energy flux from large scales to small is conserved, the fact that the
gridscale is much larger than the true dissipation scale wouldn't
matter: all the energy injected on the large scales is ultimately
dissipated by dissipation operating on fluctuations of some length scale,
and we don't much care whether that happens on scales
smaller by a factor of $10^{-2}$ or $10^{-10}$.  However, it is
possible that MHD\index{MHD} is different because it is subject
to (at least) two dissipation mechanisms, resistivity as well as
viscosity.  If the ratio between these two rates (the Prandtl
number) is far from unity, the nature of the turbulence could
be qualitatively altered, potentially in a way that influences
even behavior on the largest scales \cite{f07}.
These questions are particularly troubling in regard to jet launching
because (as we shall see later), the efficiency with which magnetic
fields generated by MHD turbulence in accretion disks can be used
to power jets may depend on the field topology, and magnetic reconnection,
which depends strongly on poorly-understood or
modeled dissipation mechanisms, alters topology.

Another problem whose origin lies in finite computational power
is the difficulty in using one simulation to predict behavior of
the system under different parameters.  In contrast to analytic
solutions, a single numerical simulation only rarely points clearly
to how the result would change if its parameters were altered.  However,
the scale of the effort required to run simulations makes
it nearly impossible to do large-scale parameter studies: Typical
computer allocations allow any one person to do at most 5--10
simulations per year; each one may take a month or so to run to
completion; and analysis of a single simulation often takes several
months of human time.  Thus, scaling the results to other circumstances
is in general very challenging.

Lastly, there is a fundamental limitation to numerical methods: the
solution they find depends on the initial and boundary conditions
chosen as well as on the equations and their dimensionless parameters.
Our goal is generally to find what Nature does, but the initial
and boundary conditions for a simulation are usually chosen on the
basis of human convenience.  This means, for example, that there
may be equilibrium solutions that are never encountered in a simulation
because the initial condition was, in some sense, ``too far away";
put in other language, the radius of convergence for the iterative
solution represented by the time-advance of the simulation may not
be large enough for the simulation to find the equilibrium.  Similarly,
when there are several stable equilibria available, any one simulation
can settle into at most one of them.  On other occasions, boundary
conditions can subtly prevent a simulation from evolving into a
configuration that Nature actually permits.  In the context of
jet maintenance, the most important of these imponderable issues
(or at least, the most important ones of which we are currently
aware) have to do with the magnetic field structure.  We know
neither the topology of the field supplied in the accretion flow,
nor do we know to what degree it has a fixed large-scale structure
imposed by conditions at very large distance from the black hole.
It is also very hard to imagine a way in which we might learn more
about either of these two questions.  Thus, the best we can do
is to explore the consequences of a variety of choices; if we
are fortunate, we may find that some of the options don't make
much difference to the astrophysical questions of greatest interest.

\section{Results}
\label{sec:results}

     As we have seen, what determines the strength and structure of the
magnetic field near the black hole is the central question for studies of
relativistic jet launching.  Any attempt to answer this question must
therefore be carefully structured so as to avoid embedding the answer
in the assumptions.  Because it is not always easy to predict which
assumptions are truly innocuous, here we will report what has been
found to date and then discuss potential dependences upon parameters,
boundary conditions, and the set of physical processes considered.

\subsection{The simplest case}

   The configuration that has been studied most is arguably the simplest:
A finite amount of mass in an axisymmetric hydrostatic equilibrium is placed
in orbit a few tens of gravitational radii from the black hole, its
equatorial plane identical to the equatorial plane of the rotating spacetime.
The initial magnetic field is entirely contained within the matter, so
that there is no net magnetic flux and no magnetic field on either the
outer boundary or the event horizon.  Because the gas density declines
monotonically away from a central peak, one
can identify the initial magnetic field lines with
the density contours, so that they form large concentric dipolar loops
\cite{mg04,dVhkh05,hk06}.

\begin{figure}
\centering
\includegraphics[height=6cm]{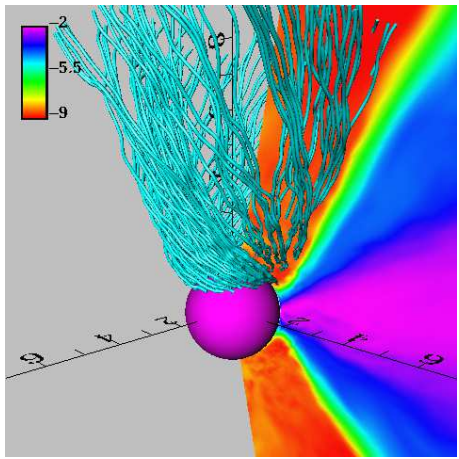}
\caption{Field lines near a black hole rotating with $a/M = 0.9$ \cite{hkh04}.
The background colors illustrate matter density on a logarithmic scale,
calibrated by a color bar found in the upper left-hand corner.  The
axes show Boyer-Lindquist radial coordinate.}
\label{fig:rotfieldlines}
\end{figure}

     Starting from this configuration, the field line segments on the
inner side of the loops are rapidly pushed toward the black hole, arriving
there well before much accretion of matter has taken place.  Unburdened by
any significant inertia, they expand rapidly into the near-vacuum above
the plunging region, where a centrifugal barrier prevents any matter with
even a small amount of angular momentum from ever entering.  Because the
vertical component of the magnetic field on the inner field lines has a
consistent sign in both hemispheres, within a short time individual
field lines run from far up along the rotation axis in the upper hemisphere
to equally far down along the axis in the lower hemisphere, passing close
outside the horizon when they cross the equatorial plane.  Just as predicted
by the Blandford-Znajek picture, a rotating black hole forces an otherwise
radial field to develop a transverse component, as shown in
Fig.~\ref{fig:rotfieldlines}.  Note how the winding of the field lines
is tightest close to the event horizon, where frame-dragging is strongest.

       We stressed earlier that the classical Blandford-Znajek model
predicts the luminosity in terms of the radial component of the magnetic
field on the horizon and the rotation rate of the field lines, but gives
no guidance about how to estimate the former, and only a guess about the latter.
In the Newtonian limit far from the innermost stable circular orbit,
the field-strength is tightly coupled to the accretion rate because
angular momentum conservation requires
\begin{equation}
- \int \, dz \, \langle B_r B_\phi \rangle \simeq {\dot M \Omega \over 2\pi}.
\end{equation}
Close to the black hole, this relation remains correct as an
order of magnitude estimator.  Consequently, when the field is
directly associated with accretion, its intensity near
the black hole should be proportional to $\dot M$.  Using simulation
data, \cite{hk06} confirmed this expected proportionality, but
found that spin matters, too:
\begin{equation}
\langle \frac{B^2}{8\pi} \rangle \simeq \frac{0.01}{1 - a/M}
                   \frac{\dot M c^5}{(GM)^2},
\end{equation}
where the magnetic energy is measured in the fluid frame at the
event horizon.

     Despite the fact that the accretion flow sets the scale of
magnetic intensity, the inertia of matter has little to do with
field dynamics near the rotation axis.  As already mentioned, any
matter with even the angular momentum of the last stable orbit
is excluded from a cone surrounding the axis.  Within that cone,
the field is force-free in the sense that $||B||^2/(4\pi \rho h c^2) \gg 1$.
However, it should also be recognized that simulations like those done
to date are not able to define quantitatively just how large this ratio
is.  Precisely because the matter's angular momentum makes the interior
of the cone forbidden territory, any matter in
the jet cone got there by some numerical artifact, either exercise
of a code density floor or through numerical error associated with
insufficient resolution of the extremely sharp density gradient
at the centrifugal barrier.  The enthalpy of the gas is equally
poorly known, but that is because the treatment of thermodynamics
in global simulations thus far is so primitive.  Thus, the most
one can say at this point is that, in a purely qualitative sense,
the interior of the jet cone should be magnetically dominated.
For exactly these reasons, the Lorentz factor of the outflow is
equally ill-determined.

\begin{figure}
\centering
\includegraphics[height=6cm,angle=90]{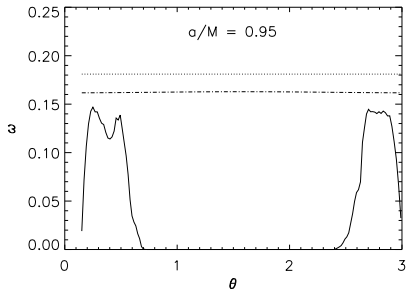}
\caption{Radially- and time-averaged local field line rotation rate (solid
curve) as a function of polar angle for a simulation in which the
black hole has spin parameter $a/M = 0.95$ \cite{hk06}.  The dotted curve
is half the black hole's rotation rate, the dash-dot curve is half
the rotation rate of the inner boundary of the simulation, which
is slightly outside the event horizon.
\label{fig:omegafield}}
\end{figure}

Although even defining a field line rotation rate for
time-dependent non-axisymmetric fields is a bit dicey, it is possible
to do so in an approximate way by monitoring the azimuthal velocity
of the matter attached to the field lines, and subtracting off the
portion attributable to sliding along the field lines.  For example,
if one defines the ``transport velocity" by $V^i \equiv u^i/u^t$, in
Boyer-Lindquist coordinates the local field line rotation rate $\omega$ can
be written as
\begin{equation}
\omega = V^\phi - B^\phi{V^r B^r g_{rr} + V^\theta B^\theta g_{\theta\theta}
     \over (B^r)^2 g_{rr} (B^\theta)^2 g_{\theta\theta}}.
\end{equation}
As Fig.~\ref{fig:omegafield} shows, after
averaging radially, the rotation rate is close to the MacDonald-Thorne guess,
perhaps 10--$20\%$ less, with surprisingly little variation in
polar angle through the jet \cite{mg04,hk06}.

   Strikingly, even when the accreting matter orbits {\it opposite}
the sense of rotation of the black hole, the field lines in the jet
rotate with the black hole, not the matter \cite{hk06}.  This fact
makes it clear that the primary motive power for the jet is drawn
from the black hole, {\it not} matter circulating deep inside the
ergosphere to which the field lines are attached.

      Normalized to the rest-mass accretion rate in traditional
fashion, the jet luminosity can be sizable when the black hole
spins rapidly.   Both \cite{mg04} and \cite{hk06} made rough
analytic fits to the dependence on spin:
\begin{equation}
\eta_{\rm EM} \simeq 
   \cases{0.068\left\{1 + [1-(a/M)^2] - 2\sqrt{1-(a/M)^2}\right\} &
               \hbox{\cite{mg04}}\cr
                0.002/(1 - |a/M|) & \hbox{\cite{hk06}}\cr}
\end{equation}
Although these expressions do not agree precisely, they agree
at a qualitative level: the efficiency can be $\sim 0.1$ at the
highest spins, but is much smaller for $|a/M| < 0.9$.  That the
efficiency can be so high when the black hole spins rapidly is
of some interest, given qualitative arguments (e.g., \cite{lop99})
that the luminosities of jets driven directly by a black hole
could never be significant.

Because the field lines in the jet rotate, it carries angular
momentum as well as energy away from the black hole.  The electromagnetic
angular momentum delivered to infinity by the jet can be comparable
to the conventional angular momentum brought to the black hole
by accretion \cite{gsm04,hk06,bhk08}, particularly when the
black hole spins rapidly.   In fact, \cite{gsm04} argued that
the electromagnetic angular momentum lost in the jet rises so
steeply with black hole spin that it may limit $a/M$ to $\simeq 0.93$,
a level considerably below the limiting spin proposed by \cite{t74},
who suggested that preferential capture of photons emitted by the
disk on retrograde orbits would cap $a/M$ at $\simeq 0.998$.

\subsection{A slightly more complicated case: the effect of field geometry}

      Beckwith et~al. \cite{bhk08} explored other options for the magnetic field's
initial geometry beyond that of the simplest model, including:
a pair of quadrupolar loops, one above and one below the plane, with
their field directions coinciding on adjacent edges; a purely toroidal
field; and a sequence of four dipolar loops, each rather narrow in radial
extent.  All of these were far less effective in terms of time-averaged
jet luminosity than the large dipolar loop initial state: the quadrupolar
case by two orders of magnitude, the toroidal by three!

    The explanation for these strong contrasts changes with the geometric
symmetry in question.  Quadrupolar loops residing in a single hemisphere
can rise buoyantly as a single unit and collapse, reconnecting with themselves.
Jets arising from that kind of field therefore tend to be highly episodic,
with only brief moments of high luminosity.  Toroidal fields cannot produce
poloidal field on scales larger than roughly the disk thickness, so their jets
are always weak.  A train of dipolar loops leads to
a succession of jet-launching and jet-destruction events, separated in
time by the difference in inflow times between their inner and outer edges.

\subsection{Open questions}

    Before directly applying the results of this simplest model
to black holes in Nature, we must first answer several questions
about its generality.

\subsubsection{Do zero net-flux simulations describe a steady-state jet?}

    Although these jets are long-lasting, they behave in certain ways
as if the flux had been placed on the black hole as an initial condition
(it arrives well before the accretion flow and then stays there, with
little change).  Moreover, although it may take a long time, eventually
the far end of the large dipole loops must reach the black hole, and at
that point reconnection will eliminate the field driving the jet.  To
gain a sense of the timescales involved, in the $\sim 10^4 GM/c^3$ duration
of these simulations, $\sim 10\%$ of the disk mass was accreted.  One
might guess, then, that the ultimate field annihilation would occur
after $\sim 10^5 GM/c^3$.  For Galactic black hole binaries, this is only
$\sim 5$~s; for AGN, it might be $\sim 5 \times 10^5$--$5 \times 10^8$~s,
or no more than a few decades.

    On the other hand, we do not know the true extent of such loops.
It is at least conceivable that they might be much larger in radial
extent, and the rapid increase of inflow time with radius might lead
to much longer intervals between jet-field destruction
events.  If so, though temporary, these jets (especially in AGN)
might be sufficiently long-lived as to be observationally interesting.
Alternatively, one of the salient empirical facts about the jets
we observe is that they are generally very unsteady.   If another
field loop follows close upon the heels of the one that just closed,
the jet might be restored equally quickly.  In this sense, the picture
just described could give a better sense of the typical state of the jet.

\subsubsection{What is the generic zero net-flux magnetic configuration?}

Different zero net-flux magnetic geometries
can lead to jets of drastically different character---how can we know
which geometry (or mix of geometries) is present in any particular
real object?  It might seem ``natural" to suppose that Nature is
messy and serves up all possibilities at once, but when the different
geometries differ in their results by orders of magnitude, the specific
proportions matter a great deal.   On the other hand, perhaps
when the source of the accretion flow is a relatively ordered structure
(e.g., a companion star in a binary system), the magnetic field delivered
to the black hole, even while possessing no net flux, might be predominantly
of a single topological character.  One might speculate (as did \cite{bhk08})
that changes in the predominant topology might be related to observable
changes in jet strength, for example, as seen in galactic black hole
binaries.

\subsubsection{What would be the effect of net magnetic flux?}

    Major open questions also remain in regard to a different sort of
ill-understood magnetic geometry: the possible presence of large-scale
magnetic field threading the accretion disk and possibly the immediate
environs of the black hole.  This is a very controversial issue:
advocates exist for almost the entire range of possible answers, from
very large to nil.  On the one hand, if the accreting gas is truly
infinitely conductive, it should hold onto any large-scale field
threading it and bring that flux to the event horizon of the black hole.
Even if successive parcels carry oppositely-directed flux, the magnitude
of the flux will have a non-zero expectation value $\sim \sqrt{N}\Phi$,
where $N$ is the number of accumulated flux ropes and $\Phi$ is their
typical individual magnitude \cite{tpm86}.  On the other hand, when
gas is accreted by a black hole, the characteristic length scale of variation
for the magnetic field contained within it must shrink by many orders of
magnitude.  Severe bends must then be created in any field lines
stretching to large distance, and extremely thorough reconnection may
therefore be expected.  In principle, {\it all} the field within the
inner part of the accretion flow may lose contact with large-scale
fields by this mechanism.  Presumably, the correct story lies somewhere
between these two extremes.

    A sense of the range of views brought to bear on this
problem may be gained by mentioning a few of the contending approaches.
\cite{lpp94} attempted to define the problem in terms of the ratio
between the accretion flow's magnetic diffusivity and effective viscosity
(in modern language, $\langle-B_r B_\phi\rangle/(6\pi \rho \Omega)$).
They argued that this ratio, an effective magnetic Prandtl number,
determines whether fields are locked to the accretion flow (and must
therefore bend sharply as they leave the disk) or can slip backwards
relative to the flow (and therefore stay pretty much where they started,
without being forced to bend substantially).
Unfortunately, the distance between this sort of ``lumped parameter"
approach and the actual microphysics is great enough that this approach
hardly suffices to decide the issue.   More recently, some (e.g. \cite{su05})
have argued that clumps of flux could self-induce inflow by losing
angular momentum through a magnetic wind, while others (e.g. \cite{rl08})
have suggested that little net magnetic flux inflow would occur because turbulent
resistivity inside the disk effectively disconnects large-scale field
lines from the accretion flow in the interior of the disk.  As this
sampling of the literature indicates, the matter remains highly unsettled.

    It is, however, potentially an important question because large-scale
flux running through the disk could have important effects both on its
accretion dynamics and on its ability to support a jet.  Simulational
studies of MHD\index{MHD} turbulence in shearing boxes suggest that net vertical
flux can strongly enhance the saturation level of the turbulence
\cite{sits04}.  How much of this effect is automatically embedded in
global simulations by field loops that connect different radii remains
unclear.  In regard to jets, a consistent sense of vertical field would
certainly serve to stabilize a base luminosity against the disruption
that quadrupolar loops, etc. can create.

     This issue could have significant observable consequences because
the specific character of the large-scale field may depend
strongly on black hole environment.  One could well imagine that
the field brought to the black hole by accretion from a stellar
companion has more organized large-scale structure than the field
threading turbulent interstellar gas accreting onto a black hole
in a galactic nucleus.  It is possible that long-term
magnetic cycles in the companion star of a binary black hole system
are connected to long-term changes in the state of the accretion flow and jet.
There could even be cases intermediate
between field entirely contained within the accretion flow and
field with large-scale constraints: this might be the situation
in a collapsing star in which extremely rapid accretion onto
a nascent black hole drives a jet that creates a $\gamma$-ray burst
\cite{bk08}.

\subsubsection{Does the condition of the matter matter?}

    As remarked earlier, neither the density nor enthalpy of the
matter in the jet can be determined quantitatively by simulations
done to date.  Although part of the difficulty is numerical, there
are also serious unsolved physics (and astrophysical contextual)
problems standing in the way.  If there were, for example, a supply
of matter with substantially smaller angular momentum than the
matter in the accretion flow proper, it would see only a low
centrifugal barrier and could enter the jet cone.  Its thermal
state would surely depend on whether there are numerous low-energy
photons passing through the jet for the gas's electrons to upscatter
(as in the case of AGN) or very few (as in the case of a microquasar
in a low-hard state).  In the sort of collapsing star that might
be the central engine for a $\gamma$-ray burst, the thermal state
of gas in the vicinity of the jet
would depend on its nuclear composition and the neutrino intensity.
Although the launching of relativistic jets by black holes may
be only weakly-dependent on the state of the matter it carries
so long as $B^2/(4\pi \rho h) \gg 1$, the subsequent dynamics of
the jet---its ultimate Lorentz factor, for example---are likely
sensitive to these considerations. 

\section{Conclusions}

Because work in this field is moving forward rapidly, any conclusions
pronounced at this stage must be limited and preliminary.  Nonetheless,
results to date are certainly strong enough to give us some confidence
that jet-launching mechanisms within the Blandford-Znajek conceptual
family play a major role in this process.  Magnetic fields that link
distant regions with regions deep in the ergosphere have now been
shown by explicit example to have the power, at least in principle,
to tap the rotational energy of spinning black holes.

The progress that has been achieved so far rests on computational
solutions of equations that, at least within the terms of the MHD approximation,
express essentially ab initio physics.  It is the direct connection
to bed-rock physics (momentum-energy conservation, Maxwell's Equations, etc.)
that gives us confidence that their results are meaningful and robust.
Greater contact with observations will become possible when the physics
contained in these calculations is expanded to include better descriptions both
of how matter enters the jet and of how it couples to radiation.
Although it will be difficult to build foundations for these parts of
the problem as securely-based as those of the dynamics, it may yet
be possible to do so at a level permitting some reasonable testing by
comparison with real data.


\end{document}